\documentclass[aps,prb,twocolumn]{revtex4}

\usepackage{graphicx}
\usepackage{dcolumn}
\usepackage{amsmath}
\usepackage{amssymb}
\usepackage{wasysym}
\usepackage{color}

\usepackage{mathtools}

\begin{document}

\title{Reply to Comment on ``Magnetotransport signatures of a single nodal electron pocket constructed from Fermi arcs''
}

\author{N.~Harrison$^1$, S.~E.~Sebastian$^2$
}

\affiliation{$^1$Mail~Stop~E536,~Los~Alamos~National Labs.,Los~Alamos,~NM~ 87545\\
$^2$Cavendish Laboratory, Cambridge University, JJ Thomson Avenue, Cambridge CB3~OHE, U.K
}
\date{\today}

\maketitle

\begin{figure}
\centering 
\includegraphics*[width=.45\textwidth]{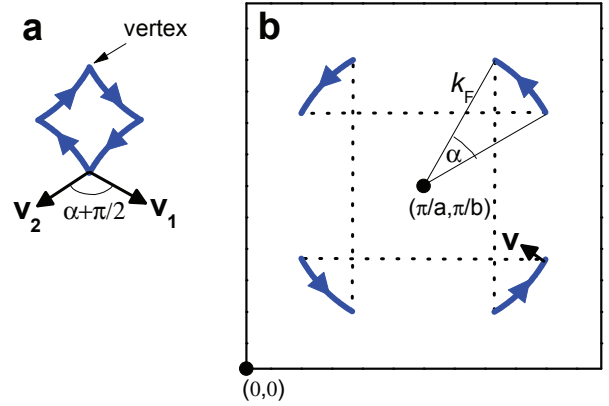}
\caption{({\bf a}), Schematic diamond-shaped electron pocket from Ref.~1, with blue arrows indicating the direction of cyclotron motion and ${\bf v}_1$ and ${\bf v}_2$ indicating the Fermi velocity direction. ({\bf b}), Schematic showing how the electron pocket is produced by connecting `arcs' of a larger hole Fermi surface, with $\alpha$ being the angle subtended by the arc and the dotted lines indicating how they are connected.}
\label{diamond}
\end{figure}

In a recent manuscript,\cite{harrison0} we showed how an electron pocket in the shape of a diamond with concave sides (see for example Fig.~\ref{diamond}a) could potentially explain changes in sign of the Hall coefficient $R_{\rm H}$ in the underdoped high-$T_{\rm c}$ cuprates as a function of magnetic field and temperature. For simplicity, this Fermi surface is assumed to be constructed from arcs of a circle connected at vertices (see Fig.~\ref{diamond}b),~\cite{harrison0} which is an idea borrowed from Banik and Overhauser.\cite{banik1} Such a diamond-shaped pocket is proposed to be the product of biaxial charge-density wave order,~\cite{harrison1} which was subsequently confirmed in x-ray scattering experiments.\cite{ghiringhelli1,chang1} Since those x-ray scattering experiments were performed, the biaxial Fermi surface reconstruction scheme has garnered widespread support in the scientific literature.\cite{maharaj1, allais1, zhang1} It has been shown to accurately account for the cross-section of the Fermi surface pocket observed in quantum oscillation measurements,\cite{harrison3,doiron2,chan0} the sign and behavior of the Hall coefficient,\cite{leboeuf1,harrison0} the size of the high magnetic field electronic contribution to the heat capacity\cite{riggs1} and more recently the form of the angle-dependent magnetoresistance.\cite{ramshaw0}

In their comment,~\cite{chakravarty0} Chakravarty and Wang raise several important questions relating to the validity of the Hall coefficient we calculated for such a diamond-shaped Fermi surface pocket. These questions concern specifically ({\it 1}~) whether a change in sign of the Hall coefficient $R_{\rm H}$ with magnetic field and temperature is dependent on a `special' form for the rounding of the vertices in Fig.~\ref{diamond}a, ({\it 2}~) whether a pocket of such a geometry can produce quantum oscillations in $R_{\rm H}$ in the absence of other Fermi surface sections and ({\it 3}~) whether a reconstructed Fermi surface consisting of a single pocket is less `natural' than one consisting of multiple pockets. Below we consider each of these in turn.

\subsubsection{Rounding of the diamond vertices}
\begin{figure}
\centering 
\includegraphics*[width=.45\textwidth]{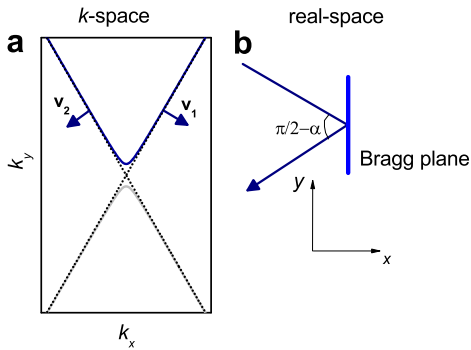}
\caption{{\bf a}, Solid lines showing the reconstructed Fermi surface in the vicinity of a vertex produced by Bragg reflection. Dotted lines indicate the Fermi surface in the absence of hybridization. ${\bf v}_1$ and ${\bf v}_2$ are velocities before and after a quasiparticle traverses the vertex. ({\bf b}) Schematic of the Bragg reflection in real-space assumed to be responsible for the sharp corner.}
\label{reflection}
\end{figure}

In our model, we assume the quasiparticle scattering rate $\tau^{-1}$ to be uniform and consider a scenario in which sharp corners on the Fermi surface are the product of electrons being Bragg reflected by the crystalline lattice, as is found to be the case in Al.\cite{banik1} In the cuprates, we assume that the sharp corners at the vertices of the diamond-shaped pocket result from the Bragg reflection of quasiparticles by the periodic potential of the charge-density wave.\cite{harrison0} While the finite size $\Delta$ of the periodic potential causes the vertices to become rounded ({\it e.g.} solid line in Fig.~\ref{reflection}a), the precise form of the rounding depends on whether the Bragg reflection is elastic or inelastic.

The mutually consistent values of the zero field Hall coefficient we obtained for the diamond using the Jones-Zener\cite{jones1} and Shockley-Chambers tube integral\cite{shockley1,chambers1} methods are neither a coincidence nor a consequence of us having assumed a special form for the rounding. Rather, they are a consequence of us having assumed the Bragg reflection to be an elastic process. In the elastic limit, the $x$ component of the velocity ${\bf v}_{\bf k}$ orthogonal to the Bragg plane reverses sign upon reflection while the $y$ component tangential to the Bragg plane remains unchanged (see Fig.~\ref{reflection}). In such a situation, the velocity of the quasiparticles evolves in a manner similar to that described by Equation (5) of Ref.~1. As long as the Bragg reflection remains an elastic process, the Hall coefficient for diamond-shaped pocket (with $\alpha>$~42.3$^\circ$) will change sign as a function of the magnetic field.

Our conclusion in Ref.~1 is in agreement with that of Banik and Overhauser,\cite{banik1} but is quite different from that reached by Ong.\cite{ong1} Ong shows that the sign of $R_{\rm H}$ does not change when a uniform $\tau^{-1}$ is replaced by a uniform magnitude $|{\bf l}({\bf k})|$ of the mean free path vector ${\bf l}({\bf k})={\bf v}_{\bf k}\tau_{\bf k}$.\cite{ong1} However, a uniform $|{\bf l}({\bf k})|$ is incompatible with elastic Bragg reflection.
To maintain a constant $|{\bf l}({\bf k})|$ while traversing the vertices, either of two unconventional scenarios would need to apply. In one scenario, the scattering rate would need to be locally suppressed at the vertices to compensate for the momentarily reduced magnitude $|{\bf v}_{\bf k}|=v_{\rm F}\cos(\alpha/2+\pi/4)$ of the velocity at the vertices given by Equation (5) of Ref.~1. In the other, the $y$-component of quasiparticle velocity would need to momentarily accelerate to a higher value at the vertices in order to maintain both ${\bf v}_{\bf k}$ and $\tau_{\bf k}$ constant. Neither of these scenarios appear to be more realistic than that we assumed in Ref.~1.

When interactions do accompany Bragg reflection, as in the case of `hot spots,' it is more likely that these will suppress the contribution to $R_{\rm H}$ from the vertices, causing a sign change in $R_{\rm H}$ to occur for smaller values of the parameter $\alpha$ in Fig.~\ref{diamond}. Possibilities include a local increase in the effective mass at the hot spots,\cite{senthil1} or an increase in the quasiparticle scattering rate.\cite{robinson1} 

\subsubsection{Oscillations in the Hall coefficient}
As Chakravarty and Wang correctly point out,\cite{chakravarty0}  the Boltzmann transport equation in the presence of Landau quantization is an intractable problem, requiring some form of approximation to be made. In our Hall effect calculations,\cite{harrison0} we chose to treat magnetic quantum oscillations in the transport using an oscillatory scattering rate $\tilde{\tau}^{-1}$. Such an approach is appealing for several reasons. First, the transport scattering rate is generally related to the number of available states for scattering in accordance with Fermi's golden rule,\cite{pippard1} causing it to approximately resemble the oscillatory electronic density-of-states. Second, the use of an oscillatory scattering rate has been shown to enable reasonably accurate modeling of quantum oscillations in the transport of several different quasi-two-dimensional metals.\cite{kikugawa1,harrison88,datars1} Third, it correctly reproduces a non-oscillatory Hall coefficient in the case of a single Fermi surface pocket of ideal circular geometry.

Chakravarty and Wang are correct in stating that there are no oscillations in the Hall coefficient of the diamond-shaped pocket in the limits $\omega_{\rm c}\tau\ll$~1 and $\omega_{\rm c}\tau\gg$~1. In fact, the oscillations of the Hall coefficient vanish in both limits $\omega_{\rm c}\tau\ll$~1 and $\omega_{\rm c}\tau\gg$~1 for a wide variety of different Fermi surface models (in the absence of a quantum Hall effect). To illustrate generality, it is instructive to consider the example of a two band metal, which, when composed of electron and hole pockets with different sizes and mobilities, yields similar magnetic field-dependent behavior to that of our diamond-shaped pocket.\cite{banik1} The vertices and concave sides of the diamond give opposing electron- and hole-like contributions to the Hall coefficient, with the concave sides dominating over the vertices in weak magnetic fields (when $\alpha>$~42.3$^\circ$). If $R_1$ and $R_2$ are the individual Hall coefficients in the two band model, then $R_{\rm H}=R_1R_2/(R_1+R_2)$ in the limit $\omega_{\rm c}\tau\rightarrow\infty$, which is non-oscillatory owing to the contributions from quantum oscillatory diagonal terms ($\sigma_{xx}$ and $\sigma_{yy}$) containing $\tau^{^-1}$ having vanished. 

However, contrary to Chakravarty and Wang,\cite{chakravarty0} we argue that magnetic quantum oscillations in the underdoped high-$T_{\rm c}$ superconductors are in fact observed in the intermediate regime in which $\omega_{\rm c}\tau\approx$~1.\cite{sebastian1,sebastian9} In such a regime, the quantum oscillatory diagonal conductivity terms containing $\tau^{-1}$ do not vanish in a two band metal, leading to quantum oscillations in $R_{\rm H}$.\cite{kikugawa1} In fact, Chakravarty has recently advocated such a scenario.\cite{chakravarty1} In a very similar way to a two band metal, the Hall coefficient of a diamond-shaped pocket also contains non-vanishing contributions from $\tau^{^-1}$ in the intermediate regime $\omega_{\rm c}\tau\approx$~1, as demonstrated algebraically by Banik and Overhauser.\cite{banik1} We therefore argue that in a very similar way to a two band metal, a diamond-shaped pocket will also exhibit quantum oscillations in $R_{\rm H}$.\cite{harrison0} 

\subsubsection{Single versus multiple pockets}
The occurrence of multiple pockets in the majority of Fermi surface reconstruction scenarios\cite{chakravarty1,maharaj1,allais1,millis1,yao1} does not make these scenarios more likely, as argued by Chakravarty and Wang.\cite{chakravarty0} Other considerations such as the small value of the electronic heat capacity at high magnetic field in fact constrain the number of pockets per CuO$_2$ plane to unity,\cite{sebastian9} making such multiple pocket scenarios less likely.
%
%
There are at least two other materials\cite{chaikin1,goddard1} in which Fermi surface reconstruction by incommensurate spin- and or charge-density wave order has been shown experimentally to yield only a single reconstructed pocket. With this in mind, two Fermi surface reconstruction scenarios based on biaxial charge-density wave order\cite{harrison2,harrison3} have been shown to be capable of producing a reconstructed Fermi surface consisting of only a single pocket. 

\subsection{Acknowledgements}
This work was supported by the US Department of Energy "Science of 100 tesla" BES program. The discussions that inspired this work took place at the Aspen Center for Physics, which is supported by the National Science Foundation grant PHY-1066293.


\begin{thebibliography}{99}

\bibitem{harrison0} N. Harrison, S.~E.~Sebastian, Phys. Rev. B {\bf 92}, 224505 (2015).

\bibitem{banik1} N.~C.~Banik, A.~W.~Overhauser, Phys. Rev. B {\bf 18}, 1521 (1978).

\bibitem{harrison1} N.~Harrison, S. E. Sebastian, Phys. Rev. Lett. {\bf 106}, 226402 (2011).

\bibitem{ghiringhelli1} G.~Ghiringhelli {\it et al.}, Science {\bf 337}, 821 (2012).

\bibitem{chang1} J.~Chang {\it et al.}, Nature Phys.{\bf 8}, 871 (2012).

\bibitem{maharaj1} A. V. Maharaj, P.~Hosur, S. Raghu, Phys. Rev. B {\bf 90}, 125108 (2014).

\bibitem{allais1} A.~Allais, D.~Chowdhury, S.~Sachdev, 
Nature Commun. DOI: 10.1038/ncomms6771 (2014).

\bibitem{zhang1} L.~Zhang, J.-W. Mei,  
preprint arXiv:1408.6592 (2014).

\bibitem{doiron2} N.~Doiron-Leyraud~{\it et al.}, Nature Commun. 
DOI: 10.1038/ncomms7034 (2015).

\bibitem{chan0} M.~K.~Chan {\it et al.}, Nature Commun. {\bf 7}, 12244 (2016).

\bibitem{harrison3} N.~Harrison, Phys. Rev. B {\bf 94}, 085129 (2016).

\bibitem{leboeuf1} D.~LeBoeuf {\it et al.}, Nature {\bf 450}, 533 (2007).

\bibitem{riggs1} S.~C.~Riggs {\it et al.}, 
Nature Phys. {\bf 7}, 332 (2011).

\bibitem{ramshaw0} B.~J.~Ramshaw {\it et al.}, preprint arXiv:1607.07145 (2016); {\it ibit.} Quantum Materials (in press 2017).

\bibitem{chakravarty0} S.~Chakravarty, Z.~Wang, preprint arXiv:1612.04870 (2016)


\bibitem{jones1} H.~Jones, C.~Zener, Proc. Roy. Soc. A {\bf 145} (1934).

\bibitem{shockley1} W.~Shockley, Phys. Rev. {\bf 79} 191 (1950).

\bibitem{chambers1} R.~G.~Chambers, Proc. Roy. Soc. London Section 
A {\bf 65}, 458 (1952).

\bibitem{ong1} N.~P.~Ong, Phys. Rev. B {\bf 43}, 193 (1991).


\bibitem{senthil1} T.~Senthil, 
preprint arXiv:1410.2096 (2014).

\bibitem{robinson1} P.~Robinson, N.~E.~Hussey, Phys. Rev. B {\bf 92}, 220501 (2015).

\bibitem{pippard1} A. B. Pippard, {\it Magnetoresistance in Metals} (Cambridge
University Press, Cambridge 1989).

\bibitem{kikugawa1} N.~Kikugawa {\it et al}, J. Phys. Soc. Japan {\bf 79}, 024704 (2010).

\bibitem{harrison88} N.~Harrison {\it et al.}, Phys. Rev. B. {\bf 54}, 9977 (1996).

\bibitem{datars1} A.~E.~Datars, J.~E.~Sipe, Phys. Rev. B {\bf 51}, 4312 (1995).

\bibitem{chakravarty1} S.~Chakravarty, H.-Y.~Kee, Proc. Natl. Acad. Sci. USA~{\bf 105}, 8835 (2008).


\bibitem{sebastian1} S. E. Sebastian, S.~E.~{\it et al.}, Nature {\bf 511}, 61-64 (2014).

\bibitem{sebastian9} S.~E.~Sebastian, N.~Harrison, G.~G.~Lonzarich, Rep. Prog. Phys. {\bf 75}, 102501 (2012).

\bibitem{millis1} A.~J.~Millis, M.~R.~Norman, Phys. Rev. B~{\bf 76}, 220503 (2007).


\bibitem{yao1} H.~Yao, D.~H.~Lee, S.~A.~Kivelson, Phys. Rev. B~{\bf 84}, 012507 (2011).

\bibitem{chaikin1} P.~M.~Chaikin, J. Phys. l France {\bf 6}, 1875 (1996).

\bibitem{goddard1} P.~A.~Goddard {\it et al.}, Synth. Metals {\bf 120}, 783 (2001).



\bibitem{harrison2} N.~Harrison, Phys. Rev. Lett. {\bf 107}, 186408 (2011).

%


%
%
%
%
%
%
%
%
%
%
%
%
%
%
%
%
%
%
%
%
%
%
%
%
%
%
%
%
%
%
%
%
%
%
%
%
%
%
%
%
%
%
%
%
%
%
%
%
%
%
%
%
%
%
%
%
%
%
%
%
%
%
%
%
%
%
%
%
%
%
%
%
%
%
%
%
%
%
%
%
%
%
%
%
%
%
%
%
%
%
%
%
%
%
%
%
%
%
%
%
%
%
%
%
%
%
%
%
%
%
%
%
%
%
%
%
%
%
%
%
%
%
%
%

%
%
%
%
%
%
%
%
%
%
%
%
%
%

%
%
%
%

%
%
%
%
%
%
%
%
%
%
%
%
%
%
%
%




%
%
%
%
%
%
%
%
%
%
%
%
%
%
%
%
%
%
%
%
%
%
%
%
%
%
%
%
%
%
%
%
%
%





%
%
%


%
%
%
%
%

%

%
%
%
%
%
%
%
%
%
%
%
%
%
%
%
%
%
%
%
%
%
%
%
%
%
%

%
%
%
%
%
%
%
%
%
%
%
%
%










\end{thebibliography}
\end{document}